\documentclass[aps,pra,twocolumn,showpacs,superscriptaddress,secnumarabicamsmath,amssymb,
nofootinbib, groupedaddress]{revtex4}  
\usepackage{graphicx}  
\usepackage{graphics}
\usepackage{dcolumn}   
\usepackage{bm}        
\usepackage{amssymb}   
\usepackage{amsmath}
\usepackage[utf8]{inputenc}
\usepackage[parfill]{parskip}
\usepackage{longtable} 
\begin{document}
\title{Google In A Photonic Lattice}
\author{Kallol Roy}
\affiliation{Indian Institute Of Science Bangalore}
\author{Ji Liu}
\affiliation{Indiana University-Purdue University, IndianaPolis}
\author{Le Luo}
\affiliation{Indiana University-Purdue University, IndianaPolis}
\author{R.Srikanth}
\affiliation{Poornaprajna Institute of Scientific Research, Bangalore}
\author{Tapan Mishra}
\affiliation{International Center For Theoretical Science, Bangalore}
\author{Bhanu Das}
\affiliation{Indian Institute For Astrophysics, Bangalore}
\author{T. Srinivas}
\affiliation{Indian Institute Of Science Bangalore}


\date{\today}

\begin{abstract}
The quantum version of Google PageRank has recently been investigated by various groups and shown to be 
quadratically faster in time than the classical PageRank algorithm. In this paper we propose the implementation
of Quantum PageRank by a stochastic quantum walk of correlated photons in a photonic waveguide lattice,
where we allow the density matrix to evolve according to the Lindblad-Kossakowski 
master equation. This yields a stationary state, 
the diagonal of whose density matrix gives the page ranking.

\end{abstract}

\maketitle

\section{\label{sec:level1}INTRODUCTION}
Much of the success of modern day internet technologies lies on the ability to search information efficiently out of myriads
of information available. Of all the search engines available, Google search most efficiently answer the queries of user by 
exploiting the hyperlink structure of the web and ranks all the web pages that contain the ``searched phrase'' according the 
importance. Google ranks the importance of a web page $P_{i}$ as $I(P_{i})$ as the sum of importance of the set of 
pages $B_{j}$ divided by the outdegree that refers to the web page $P_{i}$\cite{PaparoMartin-Delgado}
\begin{equation} 
 I(P_{i}) = \sum_{j\in B_{i}}\frac{I(P_{j})}{\text{outdeg}I(P_{j})}
\end{equation}
If $I$ is the Column vector whose $i^{th}$ row is $I(P_{i})$ the PageRank can be expressed as the power equation
\begin{equation}
 I^{k+1} = HI^{k}
\end{equation}
where $H$ is the hyperlink matrix that describes the connectivity among the web pages in internet and is defined as:\\

\begin{equation}
H_{ij} = \begin{cases} \frac{1}{\text{outdeg}P_{j}} , & \mbox{if } P_{j} \in B_{i} \\ 0, & \mbox{if } \mbox{ otherwise} \end{cases}
\end{equation}

\footnote[1]{\tt kallol@ece.iisc.ernet.in}  
\footnote[2]{\tt plarq.sic@gmail.com}  
\footnote[2]{\tt leluo@iupui.edu} 
\footnote[3]{\tt skrintha@gmail.com}
\footnote[4]{\tt  mishratapan@gmail.com}
\footnote[5]{\tt   bpdas.iia@gmail.com}               
\footnote[1]{\tt tsrinu@ece.iisc.ernet.in}

Several patching needs to be done on the sparse hyperlink matrix $H$ because of various pecularities web connection 
topology (dangling nodes), before the equation [$2$] converges to get the PageRank vector $I$. The sparse hyperlink 
matrix $H$ is modified to a primitive, irreducible and colomn-stochastic matrix known as Google Matrix for the PageRank 
vector $I$ to converge. The Google Matrix is viewed as surfing the web by a random walker who follows the web graph 
connectivity with column-stochastic matrix $E$ with probability $\alpha$ and randomly jumping to any web node with a 
probability $(1-\alpha)$ and is defined as 
\begin{equation}
 G = \alpha E + \frac{1-\alpha}{N}\bf{1}
\end{equation}

$\bf{1}$ is a matrix whose all entries are $1$ and $\alpha$ is parameter that tunes the random walker
to follow between web connections and random jumping\cite{PaparoMartin-Delgado,QuantumNavigation}. \\
We can view the Google Matrix entries $G_{ij}$ in the language of Markov Chain as the conditional probability of 
hopping from the web page $P_{j}$ to $P_{i}$ given by
\begin{equation}
 G_{ij} = \text{Pr}(X^{n+1} = P_{i}|X^{n} = P_{j}) 
\end{equation}
where $X^{n}$ is the random variable at discrete-time instant $n$
The hopping dynamics of the random walker on the web is dictated by Google Matrix and is given by the probability
vector $\textbf{p}^{t}= \{p_{i}^{t}\}$ where $p_{i}^{t}$ is the probability that the walker is present on the web node
$P_{i}$ at time $t$ which can be described as in discrete-time
\begin{equation}
p_{i}^{t} =\sum_{j=1}^{N}G_{ij}p_{j}^{t-1}
\end{equation}
and in  continuous-time version as \cite{ConnectingDiscreteContinuousWalk}
\begin{equation}
 \frac{d}{dt}p_{i} = \sum_{j=1}^{N}(G - I)p_{j}
\end{equation}
starting with initial probability vector $\textbf{p}^{0}$ at time $t=0$, $I$ is the identity matrix and $N$ is total
number of web page present in the internet.  Each component $p_{i}^{t}$ of the stationary state vector 
$\textbf{p}^{*}$ given by
\begin{equation}
\textbf{p}^{*} = \text{lim}_{t\longrightarrow \infty}\textbf{p}^{t}
\end{equation}
gives the importance of the web page $P_{i}$ \cite{QuantumNavigation,Quantum stochastic walks}

\section{\label{sec:level2}QUANTUM PAGERANK}
Quantum PageRank is defined as importance of a quantum web page $|i\rangle$ ($|i\rangle \in Z $) and is conceived as 
the fraction of total time a quantum particle(photons, atoms etc) spends on a web page as it hos through following the Google
Matrix transition probabilities $G_{ij}$ and is defined as:
\begin{equation}
 I(|i\rangle) = |\langle i|\psi\rangle|^{2}
\end{equation}
where $|\psi\rangle$ is the state of the quantum walker after it has hops through the web.
The occupation probabilities $p_{i}$ defined for classical random walks are replaced by the projection of state of 
the walker $|\psi\rangle$ on a particular node $|i\rangle$ i.e $|\langle i|\psi\rangle|$. If we can correlate any
time-independent Hamiltonian dynamics in an $N-$dimensional Hilbert space as a quantum walk of photons on an $N-$vertex 
web defined by\cite{QuantumInformationContinuousTime, UniversalQuantumComputation}
\begin{equation}
 \frac{d}{dt}\langle i|\psi\rangle = -\frac{i}{\hbar}\sum_{j=1}^{N}\langle i|H|j\rangle\langle j|\psi\rangle
\end{equation}
If we choose the Hamiltonian of the particle so that it retains the locality of the graph so that 
it becomes quantum analog of the continuous-time version of diffusion in web i.e
\begin{equation}
 H_{ij} = \langle i|H|j\rangle = (G -I)_{ij}
\end{equation}
The Hamiltonian $H$ of the quantum particle following the web is hermitian i.e $H_{ij} = H_{ji}$. This implies
\begin{equation}
 (G -I)_{ij} = (G -I)_{ji}
\end{equation}
and restricts us to undirected graph  which departs from the actual web which is a directed graph. The occupation 
probabilities $\langle i|\psi\rangle$ does not converge because of the reversible nature of quantum walk and thus making 
the notion of Quantum PageRank impossible. \\
This motivates us to explore the environmental decoherence in quantum walk and fall
in the regime of quantum stochastic walk. Quantum stochastic walk introduces both the irreversibility and coherences
that is lacking in continuous-time quantum walk and hence solves the stationarity problem of occupation probabilities 
$\langle i|\psi\rangle$ and undirectedness of web graph. We allow the $N \times N$ density matrix $\varrho$  to evolve
markovian master dynamics and the occupation probabilities will be the diagonal elements of $\varrho$ i.e
\begin{equation}
\langle i|\varrho|i\rangle = \varrho_{ii} = p_{i} =  |\langle i|\psi\rangle|^{2}
\end{equation}
The quantum master evolution of density matrix is given by 
\begin{equation}
 \frac{d\varrho}{dt}  = \pounds\varrho  =  -i(1-\epsilon)[H,\varrho] + \epsilon \sum_{ij}\gamma_{ij}(L_{ij}\varrho L_{ij}^{\dagger}- 
\frac{1}{2}\{L_{ij}^{\dagger}L_{ij},\varrho \})
\end{equation}
where  $\epsilon\in[0,1]$ and $\gamma_{ij}$ are constants.\\
Designing the Quantum PageRank through Quantum Stochastic Walk forces us to select 
\begin{equation}
 \gamma_{ij} = G_{ij}
\end{equation}
and the dissipators $L_{ij}$ are linear operators action on a finite-dimensional Banach space.
\begin{equation}
 L_{ij} = |i\rangle\langle j|
\end{equation}
The quantum master equation reaches a steady state $\varrho^{*}$ given by
\begin{equation}
 \varrho^{*} = \text{lim}_{t\longrightarrow \infty}\varrho(t) = e^{\pounds t}
\end{equation}

if the set 
of dissipators $\{L_{ij}\}$ are self-adjoint and all the operators that commutes with
$\{L_{ij}\}$ are proportional to identity. $\pounds$ can be decomposed as a sum of Jordanian form 
\begin{equation}
 S^{-1}\pounds S = \pounds^{0}\oplus\pounds^{1}\oplus.......\oplus\pounds^{k}
\end{equation}
where
\begin{equation}
 \pounds^{k} = \begin{pmatrix} \lambda_{k} & 1 & 0 & \cdots & 0 \\ 0 & \lambda_{k}  & 1 & \ddots & 0  \\ 
 \vdots & \ddots & \ddots & \ddots    & 0 \\
 \vdots & \     & \ddots  &  \ddots   & 1  \\
  0     & \cdots & \cdots &  0        & \lambda_{k}                       \end{pmatrix}
\end{equation}
The exponential term $e^{\pounds t}$ in the stationary solution of density-matrix can be written as:
\begin{equation}
 e^{\pounds t} = S(1)\oplus e^{\lambda_{1} t}e^{N_{1} t}\oplus \cdots \oplus e^{\lambda_{k} t}e^{N_{k} t}S^{-1}
\end{equation}
where $N_{k}$ are Nilpotent matrices.

\section{\label{sec:level3} IMPLEMENTATION IN A PHOTONIC LATTICE}
We propose an architecture to implement Google Quantum PageRank with a continuous-time via constant tunneling of correlated
photons in a waveguide lattice as shown in the Fig. 1\cite{QuantumWalkCorrelatedPhotons}. The low decoherence properties of 
photons and the state of the art facilities of fabrication of waveguide arrays presents a very promising candidate for the 
implementation of PageRank in this photonic lattice. The spacing between the adjacent waveguide is order of micrometers for
evanescent coupling. \\
We choose continuous-time quantum walk to implement as it gets rid of extra coin space needed for discrete-time quantum walk and this walk evolves continuously by tunneling through 
adjacent site. The spread of the photons in waveguide lattice (or the random walker web graph)  spreads ballistically rather 
than the diffusive as in classical walk is the reason for solving the pagerank quadratically faster than the google 
pagerank algorithm today.\\
Correlated photons walking through the waveguide lattice are modeled through nearest-neighbor coupling with the 
following Hamiltonian.
\begin{equation}
 H = \sum_{j=1}^{N}[\beta_{j}a_{j}^{\dagger}a_{j} + C_{j,j-1}a_{j-1}^{\dagger}a_{j} + C_{j,j+1}a_{j+1}^{\dagger}a_{j}]
\end{equation}
where $a_{j-1}^{\dagger}$ and $a_{j}$ obey Bose-Einstein distribution and act on waveguide $j$. The quantum walker moves 
through web is viewed as the photon evolves through the waveguide of length $z$ through the unitary transform
\begin{equation}
 U = e^{-iHz}
\end{equation}
The time parameter $t$ and the length $z$ are used inter-interchangeably for the photon evolution. We plug this tight-binding
Hamiltonian of $(21)$ in to $(14)$  and solves for the stationary state of $\varrho$ using various values of $\epsilon$ 
and with an initial density matrix $\varrho_{0}$ such that all it's diagonal elements are of the value $\frac{1}{N}$

\begin{figure}
\includegraphics[scale=0.4]{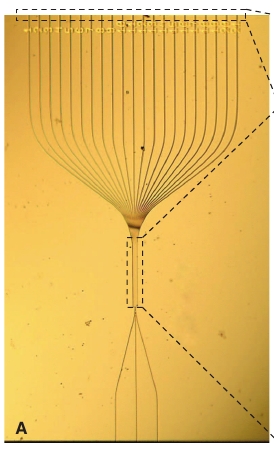}
\caption{\label{fig:epsart} Photonic Wave Guide Lattice}
\end{figure}

l

\end{document}